\documentclass[prb,12pt,a4paper,showpacs]{revtex4}
\usepackage{amssymb}
\usepackage{amsfonts}
\usepackage{amsthm}
\usepackage{amsbsy}

\theoremstyle{definition}
\newtheorem*{theor}{Theorem}

\def\R{{\mathbb{R}}}
\def\C{{\mathbb{C}}}
\def\xc{{\otimes \C}}
\def\ec{{\eta^\C}}
\def\P{{\mathbb{P}}}
\def\h{{\mathrm{hol}}}
\def\deg{{\mathrm{deg}}}
\def\d{{\bar{\partial}}}
\def\fix{{\mathrm{Fix}}}
\def\tr{{\mathrm{trace \ }}}
\def\ra{{\mathrm{rank \ }}}
\def\i{{\mathrm{i}}}

\begin{document}

\title{Geometry of Universal Magnification Invariants}

\author{M. C. Werner}
\email{mcw36@ast.cam.ac.uk}
\affiliation{Institute of Astronomy, University of Cambridge,
  Madingley Road, Cambridge CB3 0HA, United Kingdom}

\date{3 April 2009}
\pacs{02.40.Xx, 98.62.Sb}

\begin{abstract}
Recent work in gravitational lensing and catastrophe theory has shown that the 
sum of the signed magnifications of images near folds, cusps and also higher 
catastrophes is zero. Here, it is discussed how Lefschetz fixed point theory can 
be used to interpret this result geometrically. It is shown for the generic case 
as well as for elliptic and hyperbolic umbilics in gravitational lensing.
\end{abstract}
\maketitle

\section{Introduction}
This year sees the ninetieth anniversary of Eddington's eclipse
expedition to investigate the gravitational deflection of light, which
provided an early corroboration of Einstein's General
Relativity. Nowadays, gravitational lensing is an important
tool in astronomy and cosmology, and is used to address some of the current 
fundamental challenges like the properties of Dark Matter. 
A comprehensive introduction can be found, for example, in 
Schneider, Ehlers and Falco \cite{Sc92}, and Petters, Levine and 
Wambsganss \cite{Pe01}. Lensing theory is also a rich research field of its own
right within mathematical physics, in particular with applications of topological invariants and
catastrophe theory. Here, we shall revisit a recent
result concerning magnification invariants, and show that it can be understood as a
combination of those two aspects.

In the astronomically interesting limit of small deflection angles,
the physical framework for lensing is geometrical optics
subject to scalar terms of linearized General Relativity. 
Hence light rays are conserved, and the signed magnification of each lensed
image is proportional to the solid angle subtended by the ray bundle, 
taking image parity into account. Now for certain mass models acting
as lenses, one finds that the sum of the signed image
magnifications is always constant, provided the light source remains within a certain caustic domain
where the number of images is constant (and maximal), while their individual positions and magnifications, of course, are not.

This property is called a magnification invariant, and the first example was
found by Witt and Mao \cite{Wi95} for a lens consisting of two
coplanar point masses. Other examples have come to light since, 
and methods of derivation using complex analysis have been 
developed\cite{Da00, Hu01}. This type of model-dependent magnification
invariant is global in the sense that it involves the maximum number
of images produced by a lens; and that it holds in a finite domain delimited
by caustics. Furthermore, there are other magnification invariants which hold
only close to caustic singularities, but are universal in the sense that they are 
completely independent of the lens model. In fact, the universality of this type
of magnification invariant is a direct consequence of the genericity of
caustic catastrophes.

From studies of the image magnifications $\mu_i$ near folds and cusps\cite{Bl86,Sc92b,Za99}, the simplest catastrophes occurring in gravitational
lensing, it has emerged that
\[
\sum_{i=1}^n\mu_i=0
\]
for sources close to a caustic, where $n=2$ for the doublet of images due to a source near a fold, and $n=3$
for the image triplet produced by a cusp. Aazami and Petters \cite{Aa09} have
recently shown that this invariant is also true for higher caustic
catastrophes, namely for elliptic and hyperbolic umbilics both
generically and in lensing, as well as for swallowtails in the generic case. In all of these instances, the corresponding magnification invariants refer to image quadruplets, that is, $n=4$.

In the present paper, Lefschetz fixed point theory is used to interpret
this result. I would like to argue that this is interesting for two reasons. Firstly, the proof of these magnification
invariants appears to be less laborious than the one given previously 
using elementary techniques. Secondly, all magnification invariants 
considered here are analogous and differ mainly in the number of images. This
seems to hint at a more fundamental explanation beyond the details
of the individual singularities. The unified explanation given here in
terms of Lefschetz fixed point numbers appears to provide this. In
particular, it gives a more geometrical perspective on the problem,
thus tying it in with other studies of the geometry of gravitational
lensing. Since this proof relies on the Lefschetz fixed
point formula, the magnification invariants are found to be a consequence, ultimately, of the Atiyah-Bott Theorem.

The paper is structured as follows. In Section II, the relevant background in gravitational
lensing, catastrophe theory and Lefschetz fixed point theory is briefly summarized,
followed by a proof of the main result in Section III for 
generic catastrophes. More specific applications to catastrophes in gravitational 
lensing are given in Section IV, and the geometric nature of the magnification
invariants is commented on in Section V.

\section{Preliminaries}
\subsection{Gravitational lensing}
Although the proof will refer to generic maps, it may be useful to
have gravitational lensing theory in mind as a corresponding physical framework. We shall
therefore begin by outlining the mathematical structure of its
standard treatment\cite{Pe01}, adopting some suitably scaled units, and use this lensing terminology throughout.

Consider $\R^3$ with a pointlike light source at $\mathbf{y} \in S =\R^2$, the source plane,
and, in between $S$ and the observer, a mass distribution projected
into $L=\R^2$, the lens plane parallel to $S$, with surface density
$\kappa:L\rightarrow \R$. This mass acts as a gravitational lens, giving rise to a set of images seen at some
$\mathbf{x}_i\in L$ which, by Fermat's Principle, correspond to stationary points of a time
delay function, or Fermat potential, $\phi:S\times L\rightarrow \R$ such
that $\nabla\phi_\mathbf{y}(\mathbf{x}_i)=\mathbf{0}$. Now the
gravitational time delay $\psi:L\rightarrow \R$ stems from the
gravitational potential due to $\kappa$ and hence, in the given
approximation of weak fields, from a Poisson equation
$\Delta\psi=2\kappa$. Then the total time delay is given by
\[
\phi_\mathbf{y}(\mathbf{x})=\frac{1}{2}(\mathbf{x}-\mathbf{y})^2-\psi(\mathbf{x}),
\]
and the lensing map is therefore
\begin{equation}
\eta:L\rightarrow S, \ \mathbf{x}_i\mapsto\mathbf{y}, \ \mathbf{y}(\mathbf{x})=\mathbf{x}-\nabla\psi(\mathbf{x}),
\label{lensmap}
\end{equation}
the physical images being of course, mathematically speaking, the preimages of
$\eta$. Since light rays are conserved, the signed image magnification is
given by the inverse of the determinant of the Jacobian of $\eta$,
\begin{equation}
J_\eta=\left(\begin{array}{cc}
\frac{\partial y_1}{\partial x_1}|_{x_2} & \frac{\partial y_1}{\partial x_2}|_{x_1} \\ 
\frac{\partial y_2}{\partial x_1}|_{x_2} & \frac{\partial y_2}{\partial x_2}|_{x_1}
\end{array} \right), \ \mu_i=\frac{1}{\det J_\eta(\mathbf{x}_i)}.
\label{mag}
\end{equation}
For regular images, $|\mu_i|< \infty$ and $\phi$ is a Morse
function. The subset of $L$ where $\det J_\eta=0$ is called
the critical set, which maps under $\eta$ to the caustic
set in $S$. Hence the image number changes when the source crosses a
caustic, only certain types of which occur generically.

\subsection{Catastrophe theory}
These can be enumerated with reference to Thom's list of local representatives of the elementary catastrophes. The polynomial
generating potential
\[
\phi_\mathbf{u}(\mathbf{x}):\R^p\times\R^q\rightarrow \R,
\]
where $\mathbf{u}\in \R^p$ are the control parameters and
$\mathbf{x}\in\R^q$ are the state variables, gives rise to a caustic
set as discussed above, which traces out a bifurcation set or big
caustic in the space of control parameters. Since we are ulimately interested in
gravitational lensing, we shall limit the state space here to $q=2, \ \mathbf{x}=(x_1,x_2)$. Also, the
parameter space has $p\geq 2$ because of the source coordinates plus,
potentially, some physical parameters of the lens model itself. For purposes of classification, 
the angle $\beta$ between the tangent of the critical curve and the kernel of $J_\eta$ 
turns out to be important. Then the list of relevant generic catastrophes begins with 
the following\cite{Ma85,Pe01,Sc92}.

\begin{description}
\item[Fold.] Characterized by $\ra J_\eta=1, \ \det J_\eta=0$ and no zero of $\beta$. Hence, it is stable in a family of generating potentials $\phi:\R^2\times\R^2\rightarrow \R$ defined by two parameters $\mathbf{y}=(y_1,y_2)$. For given parameter values, the generic map $\eta:\R^2\rightarrow\R^2$ has up to two common solutions.
\begin{eqnarray*}
\phi_\mathbf{y}(\mathbf{x})&=&y_1x_1+y_2x_2-\frac{1}{2}x_1^2-\frac{1}{3}x_2^3, \\
\eta(\mathbf{x})&=&\left(x_1,x_2^2\right).
\end{eqnarray*}
\item[Cusp.] Characterized by $\ra J_\eta=1, \ \det J_\eta=0$ and a simple zero of $\beta$. It is therefore also stable in a generating family $\phi:\R^2\times\R^2\rightarrow \R$ defined by two parameters $\{y_1,y_2\}$ and, for given parameter values, the generic map $\eta:\R\times\R^2\rightarrow\R^2$ has up to three common solutions.
\begin{eqnarray*}
\phi_\mathbf{y}(\mathbf{x})&=&y_1x_1+y_2x_2-\frac{1}{2}x_1^2-\frac{1}{2}y_1x_2^2-\frac{1}{4}x_2^4,\\
\eta(\mathbf{x})&=&\left(x_1,x_1x_2+x_2^3\right).
\end{eqnarray*}
\item[Swallowtail.] Characterized by $\ra J_\eta=1, \ \det J_\eta=0$ and a double zero of $\beta$. Thus, it is stable in a generating family $\phi:\R^3\times\R^2\rightarrow \R$ defined by three parameters $\{c,y_1,y_2\}$. For given parameter values, the generic map $\eta_c:\R\times\R^2\rightarrow\R^2$ has up to four common solutions.
\begin{eqnarray*}
\phi_{c,\mathbf{y}}(\mathbf{x})&=&y_1x_1+y_2x_2-\frac{1}{2}y_2x_1^2-\frac{1}{2}x_2^2-\frac{1}{3}cx_1^3-\frac{1}{5}x_1^5,\\
\eta_c(\mathbf{x})&=&\left(x_1x_2+cx_1^2+x_1^4,x_2\right).
\end{eqnarray*}
\item[Elliptic Umbilic.] Characterized by $\ra J_\eta=0$, giving three equations, and an inequality. Hence it is stable in a generating family $\phi:\R^3\times\R^2\rightarrow \R$ defined by three parameters $\{c,y_1,y_2\}$. For given parameter values, the generic map $\eta_c:\R\times\R^2\rightarrow\R^2$ has up to four common solutions.
\begin{eqnarray*}
\phi_{c,\mathbf{y}}(\mathbf{x})&=&y_1x_1+y_2x_2+c(x_1^2+x_2^2)+x_1^3-3x_1x_2^2, \\ 
\eta_c(\mathbf{x})&=&\left(3x_2^2-3x_1^2-2cx_1,6x_1x_2-2cx_2\right).
\end{eqnarray*}
\item[Hyperbolic Umbilic.] Characterized by $\ra J_\eta=0$ as before, and a different inequality. Therefore, it is also stable in a generating family $\phi:\R^3\times\R^2\rightarrow \R$ defined by three parameters $\{c,y_1,y_2\}$. For given parameter values, the generic map $\eta_c:\R\times\R^2\rightarrow\R^2$ has again up to four common solutions.
\begin{eqnarray*}
\phi_{c,\mathbf{y}}(\mathbf{x})&=&y_1x_1+y_2x_2+cx_1x_2+x_1^3+x_2^3, \\
\eta_c(\mathbf{x})&=&\left(-3x_1^2-cx_2,-3x_2^2-cx_1\right).
\end{eqnarray*}
\end{description}

\subsection{Lefschetz fixed point theory}
The proof the main theorem will be based on the Lefschetz
fixed point formula. The real and complex
versions are briefly summarized here, which can be understood as special
cases of the Atiyah-Bott Theorem\cite{At67,At68}. See, for instance,
Griffiths and Harris\cite{Gr78} for a comprehensive discussion.

Let $M$ be a compact, orientable, real manifold
without boundary. Then the smooth map $f:M\rightarrow M,\ \dim_\R(M)=d,$ gives rise to a 
set of fixed points $\fix(f)=\{x\in M:f(x)=x\}$, that is,
intersections of the graph $\{(x,f(x))\}\in M \times M$ with
the diagonal $\{(x,x)\}\in M \times M$. Since the pull-back of
$f$ commutes with the exterior derivative according to
$f^\ast\circ\mathrm{d}=\mathrm{d}\circ f^\ast$, $f$ induces a map on
the de Rham cohomology classes $H^k_\mathrm{d}(M)$. Thus, Lefschetz
fixed point theory establishes a connection between local properties
of the fixed points of $f$, the fixed point indices, and the global,
topological properties of $M$. The latter is expressed by the Lefschetz
number,
\[
L(f)=\sum_{k=0}^d (-1)^k \ \tr f^\ast\left(H_\mathrm{d}^k(M)\right),
\]
a homotopy invariant, which reduces to the Euler characteristic
$\chi(M)$ if $f$ is homotopic to the identity map. The fixed point
indices are $\{+1,-1\}$ depending on the orientation of the intersection.

Next, assume that the map $f$ on a compact, complex 
manifold $M,\ \dim_\C(M)=d$ without boundary is holomorphic, that is,
$\d f=0$ in terms of the Dolbeault differential operator. In this
case, then, the crucial property is that $f^\ast$ commutes with $\d$,
thus inducing a map on the Dolbeault cohomology classes
$H_\d^{r,s}(M)$ of bidegree $(r,s)$. Hence one can define the holomorphic
Lefschetz number,
\begin{equation}
L_\h(f)=\sum_{s=0}^d (-1)^s \ \tr f^\ast \left(H_\d^{0,s}(M)\right).
\label{lef1}
\end{equation}
Provided the intersection of graph and diagonal are transversal, 
it turns out that this is connected to the local fixed point indices of $f$ via the holomorphic
Lefschetz formula,
\begin{equation}
L_\h(f)=\sum_{x \in \fix(f)}\frac{1}{\det(I_d-D_f)(x)},
\label{lef2}
\end{equation}
where $I_d$ is the $d$-dimensional identity matrix and $D_f$ is the
matrix of first derivatives with respect to local holomorphic
coordinates. The transversality condition can then be expressed as the requirement that the fixed
point indices be well defined, that is, $\det(I_d-D_f)(x)\neq 0$.

\section{Main theorem}
As indicated in the Introduction, the main result to be revisited from
a geometric point of view is a theorem about magnification invariants for generic
maps near catastrophes by Aazami and Petters\cite{Aa09}:

\begin{theor}
Let $\eta:\R^2\rightarrow \R^2$ be a generic map near
catastrophes in the sense of Section IIB, possibly dependent on control
parameters as described there. For fixed control parameters, let
$\mu_i$ be the signed magnifications of the common solutions
$\mathbf{x}_i$ of $\eta(\mathbf{x})=\mathbf{y}$ as defined in
equation (\ref{mag}). Then
\[
\sum_{i=1}^n\mu_i=0,
\]
where $n=2$ for folds; $n=3$ for cusps; $n=4$ for elliptic umbilics, 
hyperbolic umbilics and swallowtails.
\end{theor}

Since signed magnifications are non-integer numbers, it is clear that the real
version of the Lefschetz formula cannot be used to prove it. However, it has
emerged in the context of global magnification invariants that a
complexification of the lensing map using
\begin{equation}
z=x_1+\i x_2 \in L\xc=\C, \ \zeta=y_1+\i y_2 \in S\xc=\C,
\label{complex}
\end{equation}
is a fruitful approach. Then, complex magnification invariants can be derived which
hold in the real case provided that no spurious roots occur, that is,
complex roots which do not correspond to physical images\cite{Hu01}. It has previously been suggested by 
the author\cite{We07} to use this ansatz and the Lefschetz fixed point formula 
to interpret a class of global magnification invariants. But these are
model-dependent, as mentioned in the Introduction, and the technical conditions remain
to be clarified. The modified approach presented in this article is a simpler and model-independent, 
and hence more universal, application of this idea. In fact, it will also serve as an
applicaton and extension of an example of the holomorphic Lefschetz
formula discussed by Atiyah \& Bott in the original paper on their considerably more general theorem\cite{At68}.

\begin{proof}
The proof shall be divided into four steps.
\begin{description}
\item[1. Complexified map.] We begin by complexifying the map
  $\eta$ corresponding to the respective catastrophes. However,
  it is clear that the holomorphic Lefschetz formula cannot
  be applied directly to a complexification of $\eta$ by means of (\ref{complex}), for two reasons. 
Firstly, it depends on $z$ and $\bar{z}$ and is therefore not 
holomorphic, and secondly, it is not defined on a compact space
  without boundary. But we can avoid the first problem at
  the expense of the dimension, by treating $(x_1,x_2)\equiv(z_1,z_2)$ as independent
  holomorphic coordinates on $\C^2$. Then the
  corresponding map of two polynomials in two complex variables,
  \[\ec:\C^2\rightarrow\C^2,\ (z_1,z_2)\mapsto(\ec_1,\ec_2)=(y_1,y_2),\] is
  holomorphic, and each real common solution for fixed $\mathbf{y}$ is a common
  solution of $\eta(\mathbf{x})=\mathbf{y}$. Furthermore, we know from Section IIB 
that there exist domains of the parameter $\mathbf{y}$ where there are $n$ real 
common solutions of $\eta$ at finite positions in $\R^2$, for the respective 
catastrophes as given above, corresponding to the near-singular image multiplets. 
Now by considering the degrees of the polynomials, one can see from Bezout's Theorem that $n$, for the respective catastrophes, 
is also the maximum number of common solutions of $\ec$, possibly complex. Therefore, for fixed $\mathbf{y}$ in suitable domains, $\ec$ has the $n$ real common solutions of $\eta$ as stated in the theorem. Notice also that $\ec$ has no common solutions at infinity in this case.

\item[2. Fixed point map.] Next, in view of our application of
  fixed point theory, it will be useful to define a map $f:\C^2\rightarrow\C^2$ such that its set of fixed
  points $\fix(f)$ corresponds precisely to the set of common
  solutions of $\ec$. This is easily obtained,
  \begin{equation}
f(z_1,z_2)=\left(f_1(z_1,z_2),f_2(z_1,z_2)\right)=\left(z_1-\ec_1(z_1,z_2)+\zeta_1,z_2-\ec_2(z_1,z_2)+\zeta_2\right).
\label{lensmap3}
\end{equation}
  Notice also that, for fixed $(y_1,y_2)$, this implies
\begin{eqnarray*} D_f=\left(\begin{array}{cc}
1-\frac{\partial \ec_1}{\partial z_1}|_{z_2} & -\frac{\partial \ec_1}{\partial z_2}|_{z_1} \\ 
-\frac{\partial \ec_2}{\partial z_1}|_{z_2} & 1-\frac{\partial \ec_2}{\partial z_2}|_{z_1}
\end{array} \right), \end{eqnarray*}
and therefore, by definition of $\ec$ and equation (\ref{mag}),
\[
\det(I_2-D_f)=\det J_\eta=\frac{1}{\mu}.
\]
Since, by construction in the first step and for suitable domains of $\mathbf{y}$, 
the set of common solutions of $\ec$ is also the set of real common solutions of $\eta$, that is, the
images, we now have the result that the signed magnifications are in fact
\begin{equation}
\mu_i=\frac{1}{\det(I_2-D_f)(x_i)}, \ x_i\in\fix(f), \ 1 \leq i \leq n.
\label{mag3}
\end{equation}
Also, $\fix(f)$ has no fixed points at infinity in $\C^2$ in this case, 
by construction of $f$ and the first step.

\item[3. Projective fixed point map.] We now address the second
  problem mentioned in the first step, seeking a compactification that
  allows us to use the holomorphic Lefschetz fixed point formula. To
  this end, write the map $f$ of (\ref{lensmap3}) in homogeneous coordinates
  $(Z_0,Z_1,Z_2)$, where $z_1=Z_1/Z_0$ and $z_2=Z_2/Z_0$ for $Z_0\neq0$ as usual. Let
  $m=\max(\deg(\ec_1),\deg(\ec_2))$ and consider the following
  holomorphic map on complex projective space,
\begin{eqnarray*}
F&:&\C\P^2\rightarrow \C\P^2, \ \ (Z_0:Z_1:Z_2)\mapsto(F_0:F_1:F_2), \
\ \mbox{where} \\
F_0&=&Z_0^m, \\
F_1&=&Z_1 Z_0^{m-1}-Z_0^{m-\deg(\ec_1)}\ec_1(Z_0,Z_1,Z_2)+\zeta_1
Z_0^m, \\
F_2&=&Z_2 Z_0^{m-1}-Z_0^{m-\deg(\ec_2)}\ec_2(Z_0,Z_1,Z_2)+\zeta_2
Z_0^m.
\end{eqnarray*}
We need to establish that this is in fact a well-defined
map on $\C\P^2$. To see this, note first of all that $m\geq 2$ by the definiton of
$\ec$ in the first step and the list of catastrophes in Section
IIB. Now $F$ is well-defined except at any $(Z_0:Z_1:Z_2)$ where $F(Z_0:Z_1:Z_2)=(0:0:0)\neq \C\P^2$. Assume there exists such a
point. Since this implies $Z_0=0$ from the definition of $F_0$, the
respective first terms of $F_1,F_2$ vanish because $Z_0^{m-1}=0$ since
$m-1\geq 1$. The other terms of $F_1,F_2$ vanish for $Z_0=0,\ Z_1,Z_2\neq
0$ precisely if $f$ has fixed points at infinity in $\C^2$, by the definitions of
$f$; of $F$; and of the homogeneous coordinates. But by construction,
this is not case as noted in the second step. Therefore, $F$ is not
well-defined only for $(Z_0:Z_1:Z_2)=(0:0:0)\neq \C\P^2$, that is, it
is well defined everywhere on $\C\P^2$, as requied. This is a slightly extended 
version of the map constructed by Atiyah and Bott\cite{At68} which covers the case $\deg(\ec_1)=\deg(\ec_2)$.

One can now proceed with the properties of $F$. It will be useful to
consider the decomposition $\C\P^2=\C^2 \cup \C\P^1$ defined by $\C^2:
Z_0=1$ and $\C\P^1: Z_0=0$. Then the entire fixed point set of $F$ 
consists of $\fix\left(F|_{\C^2}\right) \cup \fix\left(F|_{\C\P^1}\right)=\fix(F)$. Specifically,
on $\C^2:Z_0=1$ one has $F_1=f_1,F_2=f_2$ by
(\ref{lensmap3}) and therefore $\fix\left(F|_{\C^2}\right)=\fix(f)$.

\item[4. Holomorphic Lefschetz number.] Finally, we need to determine
  the holomorphic Lefschetz number associated with $F$. It can be
  shown\cite{Gr78} that the Dolbeault cohomology
  classes of $\C\P^k$ are given by 
\[H_\d^{r,s}(\C\P^k)=\left\{\begin{array}{ll}
0 & \mbox{if $r\neq s$,}\\
\C & \mbox{if $r=s$.} \end{array} \right.\]
Recall from the definition (\ref{lef1}) that only the cohomology
classes of bidegree $(0,s)$ contribute to $L_\h$, that is, only $H_\d^{0,0}(\C\P^k)=\C$ in the case of complex projective space. 
Thus $L_\h=1$. One is now in a position to
use the holomorphic Lefschetz fixed point formula (\ref{lef2}) for $F$ on $\C\P^2$,
\begin{eqnarray}
1&=&L_\h(F)=\sum_{x \in \fix(F)}\frac{1}{\det(I_2-D_F)(x)} \nonumber \\
\mbox{}&=& \sum_{x\in \fix\left(F|_{\C^2}\right)}\frac{1}{\det(I_2-D_F)(x)}+\sum_{x \in
  \fix\left(F|_{\C\P^1}\right)}\frac{1}{\det(I_1-D_F)(x)}
\label{lef3}.
\end{eqnarray} 
By means of step 3 and equation (\ref{mag3}), we obtain
\begin{equation}
\sum_{x\in \fix\left(F|_{\C^2}\right)}\frac{1}{\det(I_2-D_F)(x)}=\sum_{x\in \fix(f)}\frac{1}{\det(I_2-D_f)(x)}=\sum_{i=1}^n \mu_i.
\label{lef4}
\end{equation}
Applying the holomorphic Lefschetz fixed point formula to the restriction of $F$ to $\C\P^1$ yields
\begin{equation}
\sum_{x\in \fix\left(F|_{\C\P^1}\right)}\frac{1}{\det(I_1-D_F)(x)}=1
\label{lef5}
\end{equation}
because $F$ is well-defined for $(0:Z_1:Z_2), \ Z_1,Z_2 \neq 0$, as discussed in the previous step, $L_\h=1$ by the same token as above, and $m\geq2$ so that the identity map is excluded (where the expression given for the fixed point index breaks down). In fact, this statement is the Rational Fixed Point Theorem for holomorphic maps on the Riemann sphere, which is important in complex dynamics, and an elementary proof can be found in Milnor\cite{Mi06}.

The result that $\sum_i\mu_i=0$ follows by substituting equations (\ref{lef4}) and (\ref{lef5}) into (\ref{lef3}). This concludes the proof of the theorem from the point of view of Lefschetz fixed point theory.
\end{description}
\end{proof}

\section{Gravitational lensing applications}
Going back to the starting point, one can now consider how this theorem for generic maps 
in catastrophe theory can also be applied to near-singular image multiplets in gravitational lensing theory. 
Here, the generating potential function $\phi$ is again the Fermat potential, and local expressions 
for $\eta$ near the canonical catastrophes can be found with suitable Taylor expansions\cite{Sc92}. 
These are related to the somewhat simpler generic equations given in Section IIA by non-linear 
coordinate transformations in general. Hence, the notional signed magnifications defined in 
equation (\ref{mag}) and used in the generic version of the theorem may be different from physical 
signed magnifications defined by a ratio of solid angles. The validity of the theorem in the generic 
case does therefore not immediately imply its validity in gravitational lensing, and it is necessary 
to examine those cases separately.

For the present purposes, we shall limit the discussion to umbilics in gravitational lensing, for which the theorem has been established\cite{Aa09}, and which also turns out to be the simplest lensing case in the framework presented here. The characteristic lensing maps can be written as follows\cite{Sc92},
\begin{description}
\item[Elliptic Umbilic in Lensing:]
$ \eta_p(\mathbf{x})=\left(x_1^2-x_2^2,-2x_1x_2+4px_2\right)$,
\item[Hyperbolic Umbilic in Lensing:] 
$ \eta_p(\mathbf{x})=\left(x_1^2+2px_2,x_2^2+2px_1\right),$
\end{description}
where $p$ is a parameter, and their properties are analogous to the ones discussed in Section IIA. 
Then it is clear that the constructions described in Section III are also possible here, and hence 
the theorem applies. Now it is important that the coordinates used in the equations above are related 
to the original lensing variables by linear coordinate transformations\cite{Sc92}, and the 
theorem is therefore also true for the physical signed magnifications.

\section{Concluding remarks}
Although it may seem unnecessary to construct a map $f$ from the 
lensing map $\eta$ such that images are the fixed points of $f$, I would like to argue that 
this fixed point treatment of gravitational lensing theory is in fact a rather natural approach, for two reasons.

Firstly, it can be seen from the definition of $\eta$ that this is 
possible on physical grounds because of the split between
the gravitational and geometrical time delay terms of the Fermat
potential, in the given limit of small deflection angles. Hence one can
naturally write a fixed point equation $\mathbf{y}+\nabla\psi(\mathbf{x})=\mathbf{x}$ from
(\ref{lensmap}). Of course, this split is not seen explicitly in the
present work because we use local forms of the lensing map near the
caustic catastrophes.

Secondly, it emerges from equation (\ref{mag3}) that the
transversality condition for the fixed points stated in Section IIC,
which is necessary for the Lefschetz formula to hold, translates into
the condition that image magnifications be finite. This is 
exactly the usual condition for images to be regular or, equivalently,
for $\phi$ to be a Morse function, as mentioned in Section IIA.

Finally, some comments about the geometry of these signed magnification invariants might be in order. 
Given a smooth surface, one can relate its topology to the number of stationary points by means of Morse Theory. 
One can also relate its topology to the integral of the Gaussian curvature by means of the Gauss-Bonnet Theorem. 
Now considering the surface given by the graph of the Fermat potential, it is interesting to note that 
signed image magnifications are, geometrically speaking, the inverse of the Gaussian curvature of this surface 
at its stationary points\cite{Aa09,Pe01}. The theorem, then, establishes a connection 
between discrete geometrical quantities summed over stationary points, rather than integrated, and an associated 
topological property expressed by the holomorphic Lefschetz number. Hence this seems 
like an intermediary, on a conceptual level, between Morse Theory and the Gauss-Bonnet Theorem.

\section*{ACKNOWLEDGMENTS}
I would like to thank the Science and Technology Facilities Council, United Kingdom, for financial support.

\end{document}